# Control of Electron Beam Using Strong Magnetic Field for Efficient Core Heating in Fast Ignition

T. Johzaki[1], T. Taguchi[2], Y. Sentoku[3], A. Sunahara[4], H. Nagatomo[5], H. Sakagami[6], K. Mima[7], S. Fujioka[5], and H. Shiraga[5]

[1]Graduate School of Engineering, Hiroshima University, 1-4-1, Kagamiyama, Higashi-Hiroshima, 739-8527, Japan

[2]Department of Electrical and Electronics Engineering, Setsunan University, 17-8 Ikedanaka-machi, Neyagawa, Osaka 572-8508, Japan

[3]Department of Physics, University of Nevada Reno, Nevada 89557, USA

[4]Institute for Laser Technology, 2-6 Yamada-oka, Suita, Osaka 565- 0871, Japan

[5]Institute of Laser Engineering, Osaka University, 2-6 Yamada-oka, Suita, Osaka 565- 0871, Japan

[6]National Institute for Fusion Science, 322-6, Oroshi-cho, Toki, GIFU, 509-5292, Japan

[7]The Graduate School for the Creation of New Photonics Industries, 1955-1 Kurematsu-cho, Nishiku, Hamamatsu 431-2102, Japan

*E-mail contact of main author: tjohzaki@hiroshima-u.ac.jp*

**Abstract**. For enhancing the core heating efficiency in electron-driven fast ignition, we proposed the fast electron beam guiding using externally applied longitudinal magnetic fields. Based on the PIC simulations for the FIREX-class experiments, we demonstrated the sufficient beam guiding performance in the collisional dense plasma by kT-class external magnetic fields for the case with moderate mirror ratio ($\leq 10$). Boring of the mirror field was found through the formation of magnetic pipe structure due to the resistive effects, which indicates a possibility of beam guiding in high mirror field for higher laser intensity and/or longer pulse duration.

"Control of Electron Beam Using Strong Magnetic Field for Efficient Core Heating in Fast Ignition"
Prepared for submission to Nuclear Fusion (IAEA FEC2014 paper)

1. Introduction

Fast Ignition Realization Experiments project (FIREX project) is now being conducted using Gekko XII + LFEX laser system at the institute of laser engineering, Osaka University [1]. The detailed analysis of recent experiments [2, 3] and the integrated simulations [4] showed very low core heating efficiency, *i.e.*, the energy coupling of the heating laser to the imploded core is ~ 1%. The main factors in preventing efficient heating are (1) too high fast electron energy and (2) too large beam divergence. The fast electron energy could be controlled by eliminating pre-plasma generation and by using heating laser with shorter wavelength. As for the beam divergence, however, it is difficult to control the angular spread of fast electrons since laser-plasma interactions are the strongly-non-linear phenomena. Instead of reducing the angular spread, some ideas of the guiding of the fast electron beam with large angular spread have been proposed, *e.g.*, the double cone [4-6] and the resistive guiding [7-13]. Those are based on using of self-generated magnetic fields. On the other hand, the guiding concept using the externally applied longitudinal magnetic fields has been proposed [14-16]. In the present paper, on the basis of two-dimensional (2D) particle-in-cell (PIC) simulations, we demonstrate the effects of external magnetic fields on the fast electron generation and transport in the fast ignition condition.

When the sufficiently strong longitudinal magnetic field is applied, the electrons' motion in the direction perpendicular to the beam propagation is limited, and they move along the magnetic field lines since the fast electrons are trapped by the magnetic fields lines. Thus, the beam guiding can be expected. To evaluate the required magnetic field strength for beam guiding, we have carried out 2D collisionless PIC simulations for the fast electron generation and propagation in a fully-ionized carbon plasma with the electron number density of 100 $n_{cr}$ ($n_{cr}$ is the laser critical density) [17]. The simulation results show that several kT fields are required for beam guiding in FIREX experiments [1], where the heating laser intensity is $10^{19}$ W/cm$^2$ ~ $10^{20}$ W/cm$^2$. In the simulations, however, the longitudinal magnetic fields were uniformly applied and the pulse duration were only 100fs. For more practical study, in the present paper, we carried out the simulations including the collisional effects and considering the converging fields and longer pulse duration. In the following, we discuss (1) the collisional effects and (2) the converging field effects based on the results of 2D collisional PIC simulations.

2. Guiding Effects in Collisional Plasmas with Uniform Magnetic Fields

To evaluate the collisional effects on beam guiding by external fields under the pico-second-pulse laser-plasma interactions, we carried out the 2D collisional PIC simulations using relativistic electro-magnetic particle in cell code PICLS [18]. The simulation condition is shown in **Fig.1**. The spatial resolution is 50 cells / μm, which is small enough for the collisional PIC simulations [19]. The escaping boundary condition is adopted both for particles and fields on both boundaries (*x* and *y*). The target is a fully-ionized solid-density carbon foil. The mass density is 1.1g/cm$^3$ and the electron number density $n_e$ is 300 $n_{cr}$. The exponential-profile pre-plasma with scale length of 3 μm is attached on the target front surface. The *p*-polarized (*i.e.*, electric fields in the *x-y* plane) laser pulse with the wavelength of $\lambda_L$ = 1 μm and the peak intensity of $I_L$ = 3x10$^{19}$ W/cm$^2$ is normally irradiated on the target surface. The transverse intensity profile is the Gaussian with a spot size of 6 μm FWHM (full width at half maximum). The laser focal plane is *x* = 8 μm that corresponding to the laser critical density plane. The laser intensity rises up to the peak value in 9 $\tau_L$, where $\tau_L$ is the laser period. After that the laser amplitude is kept constant during 300 $\tau_L$, and then is dropped to zero in 9 $\tau_L$. A typical simulation time is 400 $\tau_L$, which corresponds to about 1.3 ps for $\lambda_L$ = 1 μm. The external magnetic field is uniformly applied in *x*-direction. The simulations were carried out by varying the applied magnetic-field strength $B_{x,ext}$.

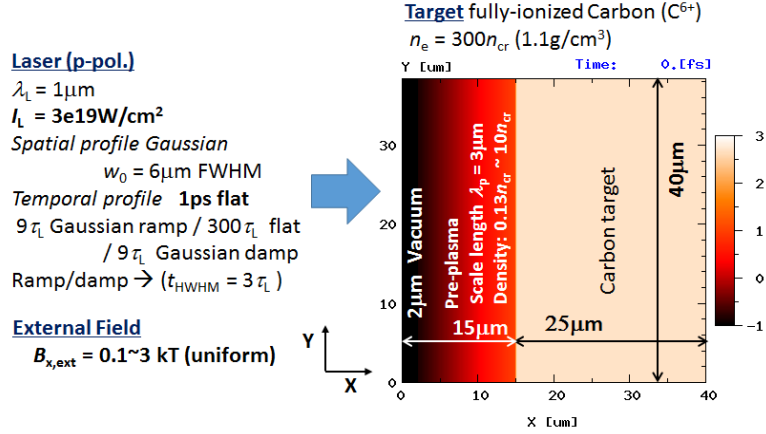

FIG.1. Simulation condition for 2D PIC simulations.

In **Fig.2**, the spatial profile of fast electron energy density and magnetic field structure are shown for $B_{x,ext}$ = 3 kT. In this case, most of the generated fast electrons are trapped by the longitudinal magnetic fields and then flow along the magnetic field lines. Then the beam does not diverge with the propagation. In addition, it is found that the longitudinal magnetic field, which is initially 3kT uniform, becomes weaker (down to less than 1 kT at the beam axis) in the central beam axis (around *y* = 20 μm). In the region around the beam edge, on the other hand, the longitudinal component is intensified (reaches ~ 4 kT). In addition, the magnetic fields in *z*-direction (~1 kT) are generated around the beam edge in a direction to confine the electron beam. So, the "magnetic pipe" [15] like structure is formed. These magnetic field evolution is basically due to the resistive effects since these structure cannot be observed in the collisionless simulations. For the discussion of B-field evolution in the dense collisional region, the simplified resistive model in widely used. In this simplified model, the temporal evolution of magnetic fields in the collisional region is described by using Faraday's law $\partial \vec{B}/\partial t = -\nabla \times \vec{E}$ and Ohm's law for the bulk plasma $\vec{E} = \eta \vec{j}_b$ ($\eta$ and $\vec{j}_b$ are the plasma resistivity and the bulk electron return current density) and by assuming the quasi current neutrality $\vec{j}_f + \vec{j}_b = 0$ ($\vec{j}_f$ is the fast electron current density);

$$\frac{\partial \vec{B}}{\partial t} = -\nabla \times \vec{E} = -\nabla \times (\eta \vec{j}_b) = \nabla \times (\eta \vec{j}_f) = \nabla \eta \times \vec{j}_f + \eta (\nabla \times \vec{j}_f). \quad (1)$$

For $B_x$ and $B_z$ components in 2D plane geometry are

$$\frac{\partial B_x}{\partial t} = -\frac{\partial E_z}{\partial y} = \frac{\partial \eta}{\partial y} j_{fz} + \eta \frac{\partial j_{fz}}{\partial y}$$

$$\frac{\partial B_z}{\partial t} = -\frac{\partial E_y}{\partial x} + \frac{\partial E_x}{\partial y} = \frac{\partial \eta}{\partial x} j_{fy} - \frac{\partial \eta}{\partial y} j_{fx} + \eta (\frac{\partial j_{fy}}{\partial x} - \frac{\partial j_{fx}}{\partial y}) \quad (2)$$

The generation of $B_z$ collimating the beam comes from $\eta (\nabla \times \vec{j}_f) \sim -\eta (\partial j_{fx}/\partial y)$ around the beam edge, as was previously reported [20-25]. The temporal evolution of $B_x$ comes from the gyration motion of fast electrons due to the longitudinal magnetic field. The gyration motion of fast electrons drives the currents in z-direction, which induces the resistive electric fields in z-direction. As the result, due to the transverse gradient of $E_z$, $B_x$ is weakened around the beam axis and is intensified around the beam edge. In the collisionless case, the reduction of $B_x$ around beam axis is observed, but smaller than that in the collisional case. The enhancement of $B_x$ and the generation of collimating $B_z$ around the beam edge are slightly observed. In the collisional plasma, thus, the fast electron beam can be guided over pico-second order time scale with self-organizing magnetic pipe structure.

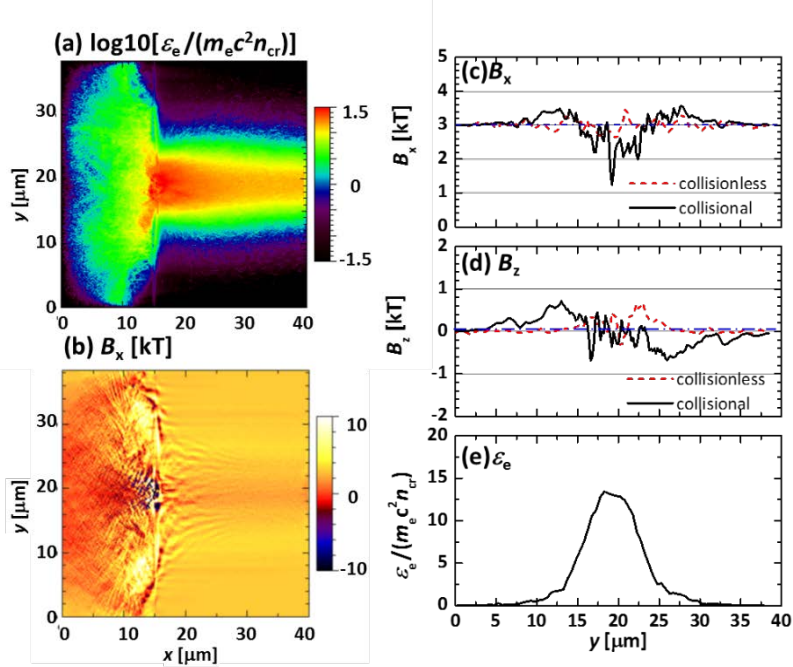

FIG.2. Spatial profiles at 990 fs for $B_{x,ext}$ = 3 kT. (a) fast electron energy density $\varepsilon_e$ normalized by $m_e c^2 n_{cr}$, where $m_e$ and c are the electron rest mass and the speed of light and (b) $B_x$ in kT and one dimensional perpendicular profiles of (c) $B_x$, (d) $B_z$ and (e) normalized $\varepsilon_e$ observed at x = 37 μm. $B_x$ and $B_z$ obtained for the collisionless case are also plotted by red – dotted lines in (c) and (d).

In Fig.3, the perpendicular distribution of beam energy density observed at 37 μm and t = 990fs are plotted for the different values of $B_{x,ext}$. The beam is diverged for the cases of $B_{x,ext}$ = 0 kT and 0.5 kT. But with increasing $B_{x,ext}$ further, the guiding effect can be observed (from $B_{x,ext}$ = 1 kT). It is found from the 2D collisional PIC simulations for ps-pulse laser plasma interactions, several-kT field is required for the electron beam guiding in FIREX class experiments where the heating laser intensity is $10^{19}$ ~ $10^{20}$ W/cm². This requirement is basically the same as the results obtained from the collisionless and short pulse simulations [17]. The generation of the required magnetic fields (several kT) have been demonstrated at experiments using a capacitor-coil target driven by ns-pulse high power laser system [26].

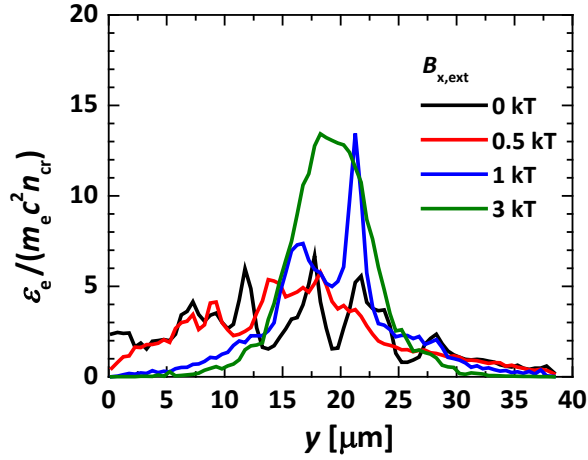

FIG.3. *Perpendicular distribution of beam energy density observed at x= 37 μm and t = 1.05 ps.*

## 3. Converging Field Effects

In FIREX experiments, we use cone-attached spherical shell targets. To apply kT-class magnetic fields, a capacitor-coil target is installed to the fusion target. Basically B-fields are applied before the fuel implosion since the timescale of B-field generation and its penetration (or diffusion) into conductive plasmas is comparable to the implosion timescale. The pre-applied magnetic fields are compressed by the fuel shell implosion, and then the field structure becomes a converging mirror one; the fields are weak around the fast electron generation point, that is the low-density relativistic laser plasma interaction region, and then it is intensified toward the dense core region. In such a converging structure, we are concerned about the reflection of fast electrons due to mirror fields. To evaluate the mirror effects, we assumed the converging fields as the initial B-field structure in PIC simulations, where the target and the laser parameters are the same as before. For setting the initial profile of converging B-field, the four linear currents flowing in z-direction were assumed and the B-field profile was calculated by the Biot–Savart law. Then, $\nabla \cdot \vec{B} = 0$ is satisfied for all cases. In **Fig.4**, an example of initial B-field structure is shown. In this case, the positions ($x$ [μm], $y$[μm]) of four currents are (61.4, 15), (61.4, -15), (-22.6, 15), (-22.6, -5) and the absolute value of current is $4 \times 10^5$ A. A part of the calculated B-field profile was used as the initial B-field profile in PIC simulations. By varying the positions and strengths of four currents, we change the mirror ratio defined as the ratio of the B-field strength on $x = 8$ μm line (the initial laser critical density line) to that on $x = 37$ μm line (fast electron observation line). In the case of Fig.4, the mirror ratio along the central laser axis $R_{M,c}$ is $B_{c,o}/B_{c,i} = 3.93$, and that along the $B$-field line passing through the point 10 μm away from the central laser axis in $y$-direction on the critical density line (corresponding to the beam edge region in Fig.2), the mirror ratio $R_{M,10\,\mu m}$ is $B_{10\mu m,o} / B_{10\mu m,i} = 4.4$. In the following discussion, we use $R_{M,10\mu m}$ as the mirror ratio. We carried out the 2D PIC simulations under the converging fields with mirror ratio of $R_{M,10\mu m} = 4.4 \sim 19.5$).

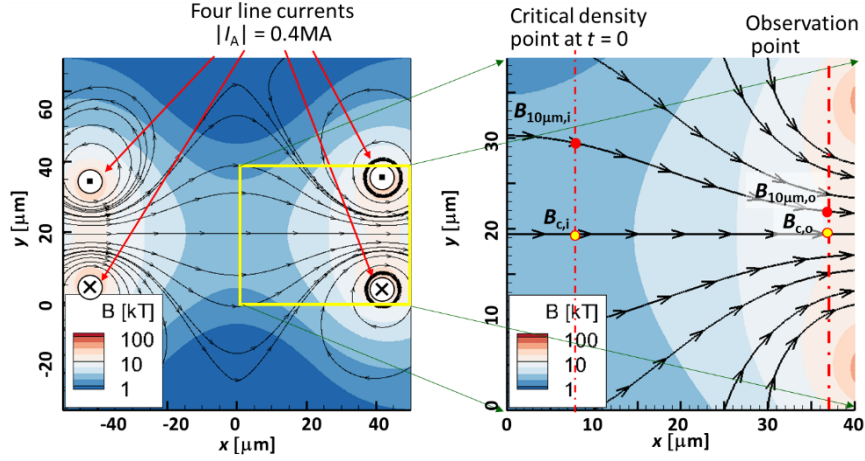

FIG.4. *An example of the initial converging B-field profile. The left figure shows positions and directions of the four linear currents flowing in z direction. The B-field profile formed by four currents was calculated by the Biot–Savart law. The right one is a part of B-field profile used as the initial profile in a PIC simulation. The mirror ratio $R_{M,C}$ is defined as $B_{c,o}/B_{c,i}$, and $R_{M, 10\mu m}$ is defined as $B_{10\mu m,o}/B_{10\mu m,i}$.*

In **Fig.5**, the spatial profile of fast electron energy density and magnetic field structure are shown for $R_{M,10\mu m} = 4.4$. In this case, part of the fast electrons are scattered or reflected due to the mirror fields. But it is found that the beam can be focused around the beam axis along the magnetic field lines. The magnetic pipe like structure is also observed. Compared with the uniform 3kT field case (Fig.2 (e)), the beam energy density around beam axis at $x = 37$ μm becomes higher (Fig.5 (e)), which means that the beam collimation is more effective in the moderate converging field case.

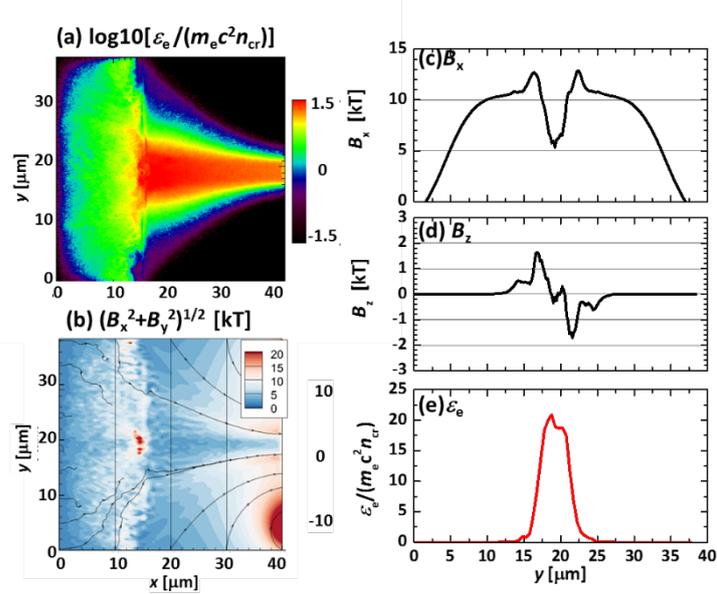

FIG.5. *Spatial profiles at 990 fs for case 1 ($R_{M,10\mu m} = 4.4$). Two dimensional profiles of (a) fast electron energy density $\varepsilon_e$ normalized by $m_e c^2 n_{cr}$, $m_e$ and c are the electron rest mass and the speed of light and (b) $(B_x^2 + B_y^2)^{1/2}$ in kT and one dimensional perpendicular profiles of (c) $B_x$, (d) $B_z$ and (e) normalized $\varepsilon_e$ observed at $x = 37$ μm.*

In **Fig.6** we compare the perpendicular distribution of the time-integrated beam energy per unit area passing through $x = 37$μm among the no-external field case, the uniform 3kT field case and the converging field cases. Compared with the uniform 3kT case, the peak value of beam energy is intensified and width of distribution becomes smaller for the converging mirror field cases. We evaluated the energy conversion efficiency $\eta_{L \rightarrow fe, 4\mu m}$ of the heating laser to the beam energy of forward-directed fast electrons observed at $x = 37$ μm (about 20 μm propagation from the interaction region) within $y = 17$ μm ~ 21 μm (4 μm width). In FIREX experiments, the imploded core size is smaller than that of laser spot. We hence set a narrower observation width than the laser spot size (6μm FWHM) in the present simulations. The energy coupling $\eta_{L \rightarrow fe, 4\mu m}$ is plotted as a function of mirror ratio $R_{M, 10\mu m}$ in **Fig.7**. The value of $\eta_{L \rightarrow fe, 4\mu m}$ is twice larger for the uniform 3 kT field case compared with the case without external fields. For the moderately-converging field cases ($R_{M, 10\mu m}$ = 4.4 and 8.2), further enhancements are observed; it increases about threefold. However, for $R_{M, 10\mu m}$ = 19.5, though the peak value of beam energy in Fig.6 is higher than that for the uniform 3 kT case, $\eta_{L \rightarrow fe, 4\mu m}$ becomes smaller than that for uniform 3 kT case and close to the value for the case without external field. The results indicate that the moderately-converging field ($R_{M, 10\mu m}$ <~ 10) can effectively guide and focus the fast electron beam generated by pico-second order relativistic laser plasma interaction. For $R_{M, 10\mu m}$ > 20, however, the mirror reflection effect exceeds the guiding effect.

We also evaluated the energy spectra of forward-directed fast electron beam ($E > 100$ keV) by counting the forward-directed fast electrons passing though $x = 37$ μm line within 4 μm width (17.4 μm < $y$ < 21.4 μm) from $t = 0$ to the end of the simulation, and show them in **Fig.8**. In all cases, the spectrum can be fitted by sum of two exponential profiles, i.e., $f(E) \propto A_L \exp(-E/T_L) + A_H \exp(-E/T_H)$. The low energy component ($E$ ~< 2MeV) has the temperature lower than 1 MeV, and the higher one ($E$ ~>5 MeV) has the temperature ~ 3 MeV. Such two temperatures profile is typically observed for the case existing pre-plasmas. Compared to the no-external field case, number of the fast electrons are increased in all energy region for the cases of uniform 3 kT field and of converging field with $R_{M, 10\mu m}$ = 4.4. With increasing mirror ratio from $R_{M, 10\mu m}$ = 4.4, the fast electron number starts to decrease, and the reduction is pronounced in the low energy region ($E \leq 1$ MeV). These low energy fast electrons mainly contribute to the core heating. Thus, the reduction of core heating efficiency is a concern even if beam energy is higher than that for the case without external field. From these simulation results, for guiding and collimating the diverging beam, the mirror ratio should be smaller than 20 at least for the pulse condition assumed here ($I_L = 3 \times 10^{19}$ W/cm$^2$ and $\tau_L = 1$ ps), and then the control of the magnetic field structure at the heating pulse injection is indispensable.

In the previous review paper [16], the beam guiding by magnetic-pipe was discussed for the ignition-scale electron beam (the beam energy $E_b$ > 100 kJ, the intensity $I_b$ ~ $10^{20}$ -$10^{21}$W/cm$^2$, the pulse duration $\tau_b$ ~ 20ps). It was reported that in the case of such high energy beams, the direction of longitudinal B-field affects the guiding performance. Though we have performed the simulations by changing the signs of $B_x$ and $B_y$ for external fields to check the dependence of guiding effect on the direction of B-fields, there was no difference on the guiding performance. This is due to lower intensity and shorter pulse duration of electron beams compared with the previous work. Our beam parameters ($E_b$ <~ 1 kJ, $I_b$ ~ $10^{19}$W/cm$^2$, $\tau_b$ ~ 1ps) are much lower than those for the ignition-scale study. In the previous study, the difference can be observed in the later phase ($t > 2$ps), but before that ($t < 2$ps) the difference is negligible. (see right figure of Fig.27 [16].) Our simulation region corresponds to the early phase of beam injection in ignition-scale simulation, and then the difference in beam propagation due to the direction of $B$-fields was not observed in our simulations.

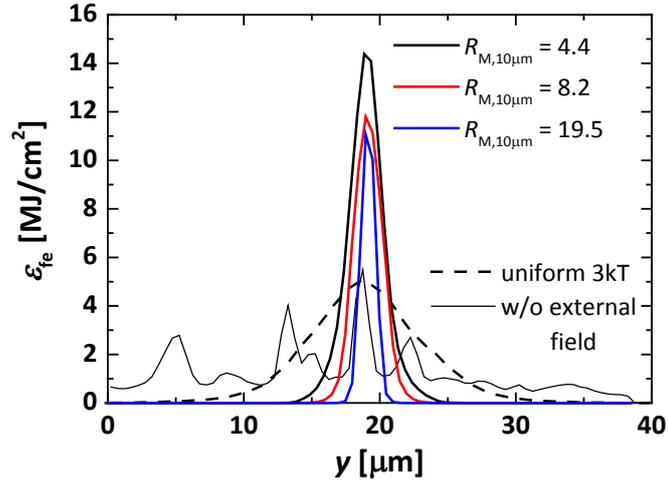

FIG.6. Comparison of perpendicular distributions of time-integrated beam energy passing through $x = 37$ μm per unit area, $\varepsilon_{fe}$, among no external field case, uniform 3 kT case and converging field cases.

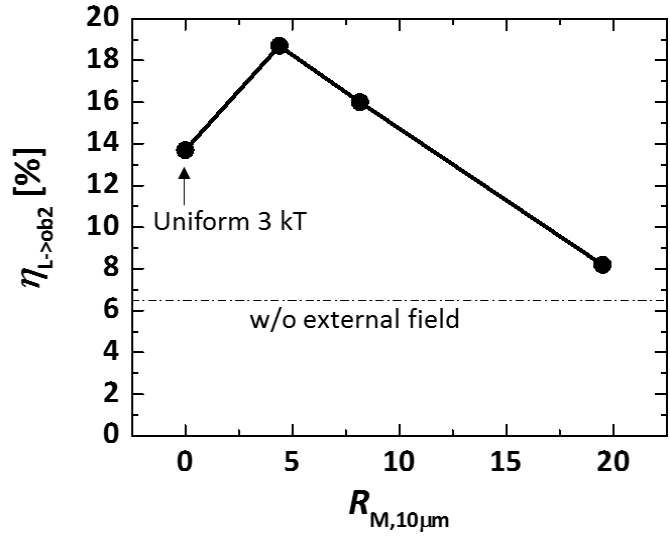

FIG.7 Energy conversion efficiency of the heating laser to the forward-directed fast electrons observed at $x = 37$ μm within $y = 17$ μm ~ 21 μm (4 μm width) as a function of mirror ratio $R_{M,10\mu m}$. The value for the case without external field is shown by a broken line.

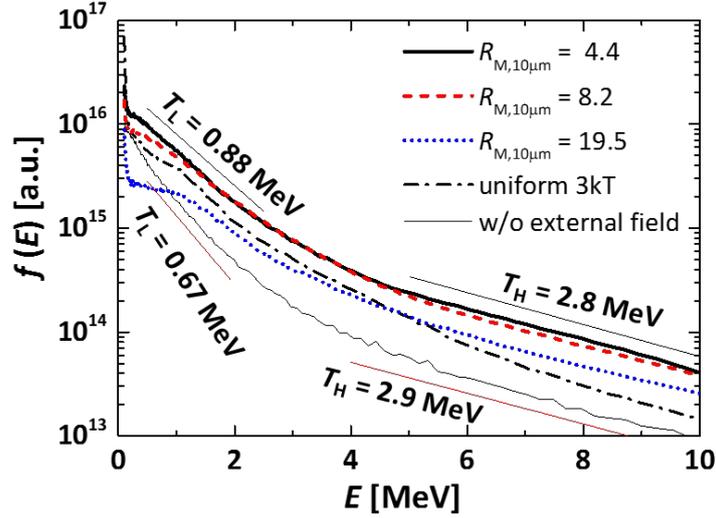

*FIG.8. Comparison of time-integrated energy spectrum of forward-directed fast electron beam (E >100 keV) observed at x = 37 μm within 4 μm width (17.4 μm < y < 21.4 μm) among no external field case, uniform 3 kT case and converging field cases.*

## 4. Concluding Remarks

The beam guiding by external fields for FIREX-class experiments was investigated by 2D collisional PIC simulations. It was demonstrated that the fast electron beam with a large divergence angle generated by a relativistic laser plasma interaction with the laser intensity of $3 \times 10^{19}$ W/cm$^2$ and the duration of 1ps can successfully focused by the moderately-converging fields ($R_{M,10\mu m} \lesssim 10$). In addition, the formation of magnetic-pipe like structure due to the resistive effects was shown, which cannot be observed in the collisionless simulations. The formation of magnetic-pipe like structure indicates a possibility of the beam collimation under the higher mirror fields which may be formed in the ignition-scale or high-gain target implosion. However, when the mirror ratio is extremely high, such as $R_M > 100$, the most of fast electrons are reflected before reaching such a high mirror ratio region, and then the formation of magnetic-pipe like structure cannot be expected. In the implosion simulation [27] for a cone-attached CD shell target with B-field externally applied by a one-turn coil, the ratio of B-field strength at the cone tip to that in the region between cone tip and dense core reaches > 100 at the maximum compression. In such a case, the fast electrons could not penetrate the very high $R_M$ region, and then they never reach the dense core. Thus, for enhancing the heating efficiency by external B-field, the optimization of applied timing and initial configuration of external B-field to form the moderate-$R_M$ B-field structure at the maximum compression. Further investigation for the beam guiding and enhancing the heating efficiency is needed, such as laser intensity dependence, pulse duration dependence and guiding under higher mirror fields including the core heating process. Also, the experimental demonstration is indispensable.

**Acknowledgements**

This work is partially conducted under the joint research projects of the institute of laser engineering, Osaka University (FIREX-project), and with the supports of the NIFS Collaboration Research program (NIFS12KUGK057 and NIFS14KNSS054) and JSPS KAKENHI Grant Number 25400534.

**Figure Captions**

*FIG.1. Simulation condition for 2D PIC simulations.*

*FIG.2. Spatial profiles at 990 fs for $B_{x,ext}$ = 3 kT. (a) fast electron energy density $\varepsilon_e$ normalized by $m_e c^2 n_{cr}$, where $m_e$ and $c$ are the electron rest mass and the speed of light and (b) $B_x$ in kT and one dimensional perpendicular profiles of (c) $B_x$, (d) $B_z$ and (e) normalized $\varepsilon_e$ observed at x = 37 μm. $B_x$ and $B_z$ obtained for the collisionless case are also plotted by red – dotted lines in (c) and (d).*

*FIG.3. Perpendicular distribution of beam energy density observed at x= 37 μm and t = 1.05 ps.*

*FIG.4. An example of the initial converging B-field profile. The left figure shows positions and directions of the four linear currents flowing in z direction. The B-field profile formed by four currents was calculated by the Biot–Savart law. The right one is a part of B-field profile used as the initial profile in a PIC simulation. The mirror ratio $R_{M,C}$ is defined as $B_{c,o}/B_{c,i}$, and $R_{M, 10μm}$ is defined as $B_{10μm,o}/B_{10μm,i}$.*

*FIG.5. Spatial profiles at 990 fs for case 1 ($R_{M,10μm}$ = 4.4). Two dimensional profiles of (a) fast electron energy density $\varepsilon_e$ normalized by $m_e c^2 n_{cr}$, $m_e$ and $c$ are the electron rest mass and the speed of light and (b) $(B_x^2 + B_y^2)^{1/2}$ in kT and one dimensional perpendicular profiles of (c) $B_x$, (d) $B_z$ and (e) normalized $\varepsilon_e$ observed at x = 37μm.*

*FIG.6. Comparison of perpendicular distributions of time-integrated beam energy flux observed at x = 37 μm per unit area, $\varepsilon_{fe}$, among no external field case, uniform 3 kT case and converging field cases.*

*FIG.7. Energy conversion efficiency of the heating laser to the forward-directed fast electrons observed at x = 37 μm within y = 17 μm ~ 21 μm (4 μm width) as a function of mirror ratio. The value for the case without external field is shown by a broken line.*

*FIG.8. Comparison of time-integrated energy spectrum of forward-directed fast electron beam (E >100 keV) observed at x = 37 μm within 4 μm width (17.4 μm < y < 21.4 μm) among no external field case, uniform 3 kT case and converging field cases.*